\documentclass[aps,prl,twocolumn,showpacs,showkeys,groupedaddress]{revtex4-1}

\usepackage{amsmath,amssymb,float,graphicx}
\usepackage[caption=false]{subfig}

\begin{document}

\title{Hybrid photonic circuit for multiplexed heralded single photons}

\author{Thomas Meany$^1$}
\email{Corresponding author: thomas.meany@mq.edu.au}
\author{Lutfi A. Ngah$^2$}
\author{Matthew J. Collins$^3$}
\author{Alex S. Clark$^3$}
\author{Robert J. Williams$^1$}
\author{Benjamin J. Eggleton$^3$}
\author{M. J. Steel$^1$}
\author{Michael J. Withford$^1$}
\author{Olivier Alibart$^2$}
\author{S\'{e}bastien Tanzilli$^2$}

\affiliation{\,\\\,\\\,\\1. Centre for Ultrahigh bandwidth Devices for Optical Systems (CUDOS), MQ Photonics Research Centre,\\ Department of Physics and Astronomy, Macquarie University, North Ryde, 2109 NSW, Australia\\
2. Universit\'{e} Nice Sophia Antipolis, Laboratoire de Physique de la Mati\`{e}re Condens\'{e}e, CNRS UMR 7336, Parc Valrose, 06108 Nice, France\\
3. CUDOS, Institute of Photonics and Optical Science (IPOS), School of Physics, University of Sydney, 2006 NSW, Australia}

\begin{abstract}
A key resource for quantum optics experiments is an on-demand source of single and multiple photon states at telecommunication wavelengths. This letter presents a heralded single photon source based on a hybrid technology approach, combining high efficiency periodically poled lithium niobate waveguides, low-loss laser inscribed circuits, and fast ($>$1~MHz) fibre coupled electro-optic switches. Hybrid interfacing different platforms is a promising route to exploiting the advantages of existing technology and has permitted the demonstration of the multiplexing of four identical sources of single photons to one output. Since this is an integrated technology, it provides scalability and can immediately leverage any improvements in transmission, detection and photon production efficiencies.
\end{abstract}


\maketitle

\begin{figure}[h]
\subfloat[]{\label{fig:4_photon_source}\includegraphics[width=0.5\textwidth]{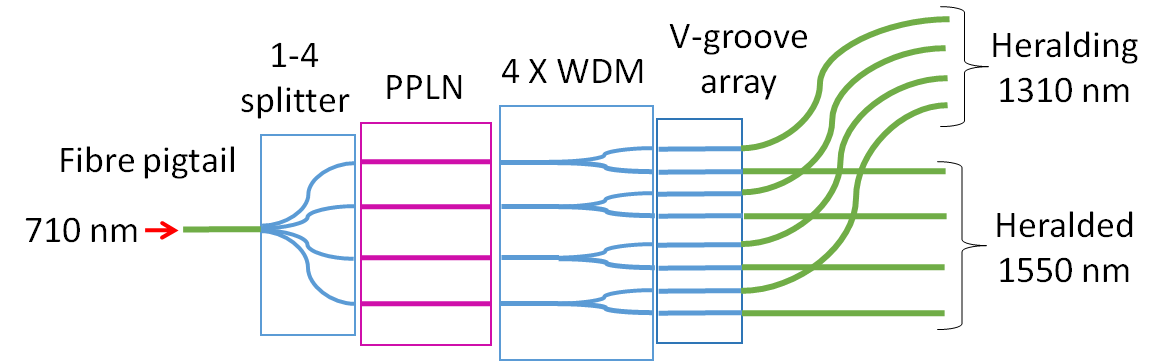}}\\
\subfloat[]{\label{fig:multiplexing_config}\includegraphics[width=0.43\textwidth]{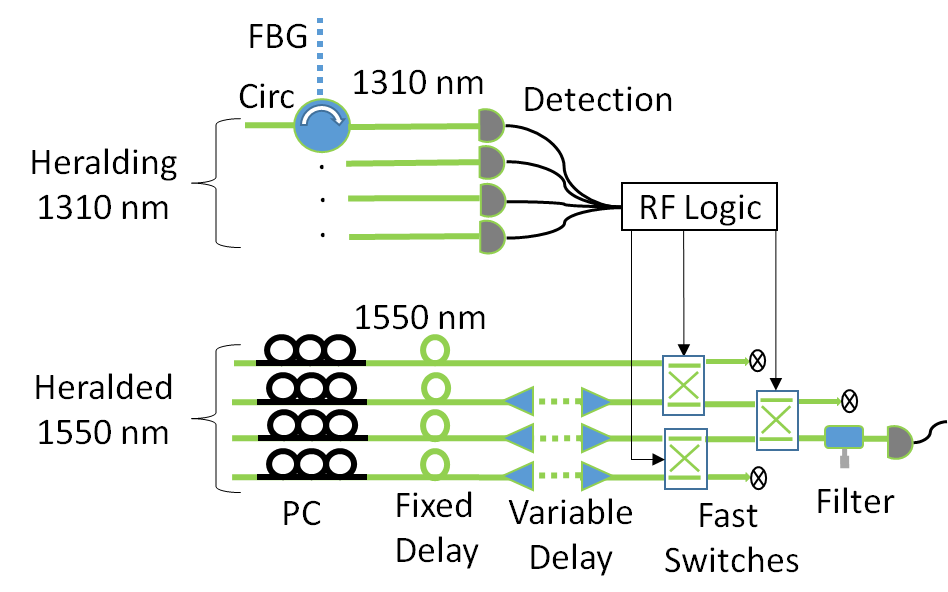}}\\
\caption{(a) A picosecond laser pumped array of 4 identical PPLN waveguides, where a pair of FLDW chips couple light in and out of the device. (b) The output of the WDMs is sent to a multiplexing setup. Four single photon channels at 1310~nm, which are filtered by fibre Bragg gratings (FBG) and circulators (Circ), herald the arrival of 1550~nm photons. Radio frequency (RF) electronics control the switches for routing 1550~nm photons and polarisation controllers (PC) control the polarisation.\label{fig:Multiplexing_exp}}
\end{figure}

Spontaneous parametric downconversion (SPDC) is one of the most commonly exploited methods to generate photons in quantum information science (QIS) applications~\cite{Kwiat1999}. However, the spontaneous nature of the process, bounding the probability that $N$ sources each simultaneously create a pair of photons, prevents the demonstration of advanced QIS protocols~\cite{Yao2012}. Furthermore, the power of the pump laser cannot simply be scaled to increase the pair generation rate since multi-photon events result in noise. Although schemes have been developed which can tolerate multi-photon pairs~\cite{Matthews2013}, the majority of protocols have stringent limitations on additional noise~\cite{Smirr2011}. 
Increasing the repetition rate of the pump laser increases the number of photon pairs generated by a source per second~\cite{Broome:11}, while the probability of generating a photon pair per pulse remains constant. In contrast, active temporal multiplexing schemes can increase the probability of obtaining a photon pair per pulse~\cite{Mower2011}, however due to implementation challenges there has been to date no experimental demonstration. To address this, Migdall \emph{et al.} proposed a spatial multiplexing scheme to increase photon rates while maintaining the corresponding noise level~\cite{Migdall2002}. This scheme relies on spatially separated heralded sources being actively routed to a single output. A bulk optic demonstration of a multiplexed source used a pair of SPDC crystals\cite{Ma2011b}. This method suffered from large space and stability requirements and was not scalable beyond a pair of sources. Integrated optics offers a route to miniaturisation, stability and scalability which enabled a number of QIS circuit demonstrations such as a controlled-NOT operation~\cite{Politi2008}, heralded multi-photon entanglement~\cite{Matthews2011} and on-chip quantum relay operation~\cite{Martin2012}. Here we propose a hybrid photon source which uses SPDC waveguides and laser written components to produce four heralded photons (Fig.~\ref{fig:4_photon_source}). We then use fast fibre-based switches to actively route photons to the desired output (Fig.~\ref{fig:multiplexing_config}). Using this architecture we have control over both the photon wavelength and number through temperature tuning and electronically addressed switches.
 
The first demonstration of integrated spatial multiplexing, using a pair of $\chi^{(3)}$ silicon photonic crystal (SiPC) waveguide sources and a single switch, has shown that this method can enhance photon rates~\cite{Collins2013}. However, the fabrication of identical on-chip SiPC sources remains an obstacle that has limited investigations of $N>2$ sources.
SPDC in $\chi^{(2)}$ nonlinear waveguide crystals, permits low-noise photon pair generation and soft-proton exchanged (SPE) waveguides, formed in periodically poled lithium niobate (PPLN), have the highest reported brightness to date~\cite{Tanz2002}. 
However, for functional monolithic circuits, nonlinear chip sections for photon pair generation and linear regions for photon routing must be combined~\cite{Martin2012,Silberhorn2013}. Incorporating a range of optical functions results in challenging fabrication procedures which require a range of separate masks.
Recently, the femtosecond laser direct-write (FLDW) technique has attracted interest for the fabrication of low-loss, linear waveguide circuits. Requiring no lithographic masks, it allows high speed device fabrication and offers an inherently 3D routing capability.
Low-loss linear complex waveguide circuits have been demonstrated in borosilicate glasses~\cite{Meany_13}. To date however, FLDW waveguides in lithium niobate have shown high loss modifications due to the combination of high refractive index and tightly packed crystalline structure~\cite{Thomas2011}. 
This motivates our goal of combining low loss linear optical circuits in borosilicate glasses with the high efficiency nonlinear properties of PPLN proton exchanged waveguides.

The layout of the on-chip portion of the experiment is shown in Fig.~\ref{fig:4_photon_source}.
The pump laser (spectrally filtered Chameleon, Coherent) provides 1.2\,ps-duration, time-bandwidth limited ($\Delta \lambda_{p}=$ 0.5\,nm) pulses, at a wavelength of $\lambda_{p}=$ 710\,nm and at a repetition rate of 76\,MHz. The pulses are sent to a polarisation maintaining fiber pigtailed to an FLDW 1-4 waveguide splitter.

\begin{figure}[h]
\includegraphics[width=0.45\textwidth]{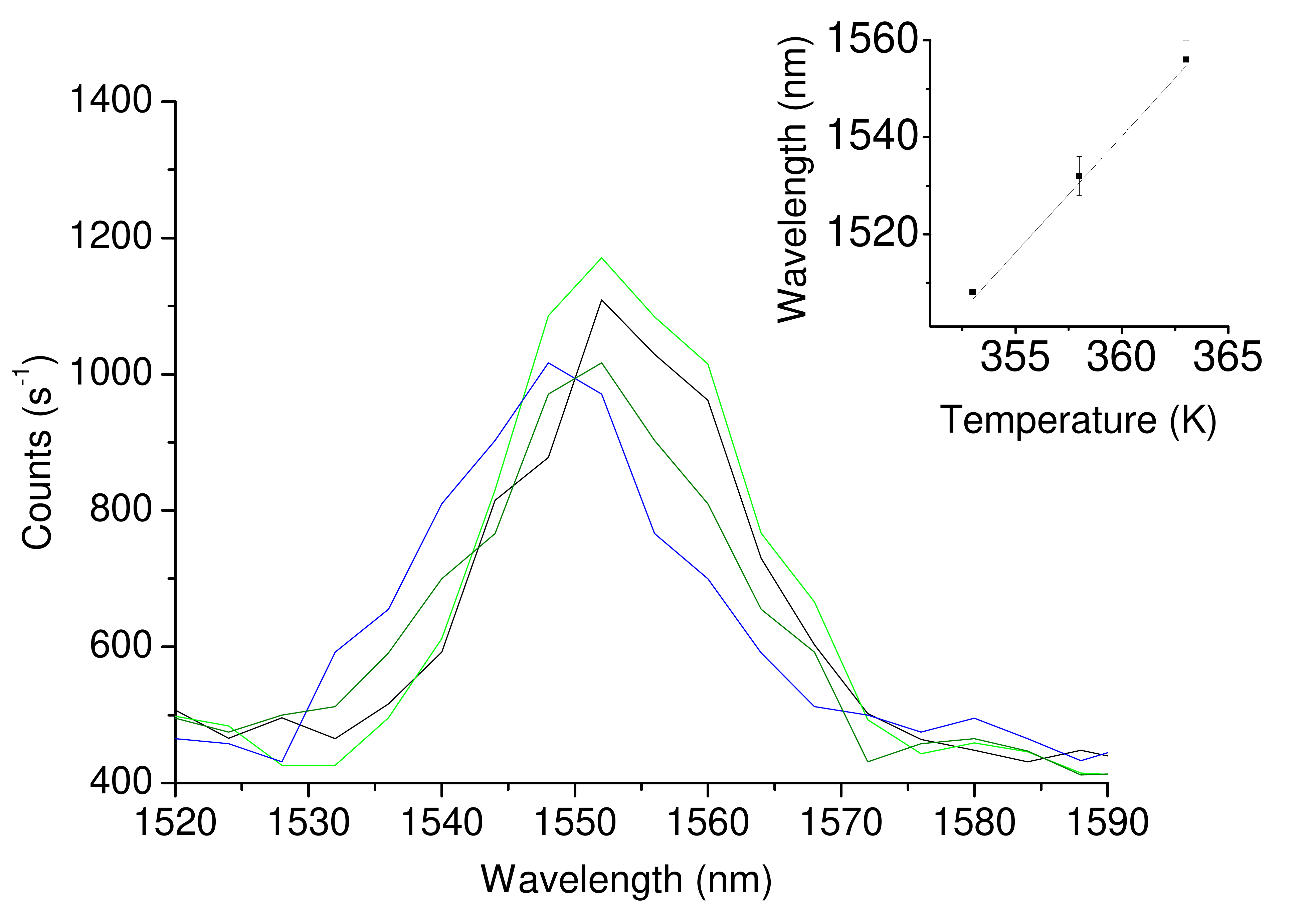}
\caption{The output spectra of the individual PPLN waveguide channels. Considering the bandwidth of the pump pulse and the length of the sample, the expected output bandwidth is 30~nm centred at 1550~nm. The colours correspond to individual PPLN channels 1 (black), 2 (green), 3 (dark green) and 4 (blue). The inset shows the change in central wavelength as a function of temperature.\label{fig:source_spectrum}}
\end{figure} 

The 1-4 splitter uses four evanescently coupled waveguides, in a 3D interaction region, to separate the input pump field~\cite{Meany_multiport}. It is then butt coupled to the input facet of the nonlinear chip to pump the four identical photon pair generators simultaneously. The latter consists of four identical quasi-phase-matched PPLN waveguides, 10\,mm in length, fabricated using SPE~\cite{Tanz2002}. They are designed to produce photon pairs, by SPDC, at 1310\,nm and 1550\,nm within a bandwidth of 30\,nm when pumped with 710\,nm picosecond pulses. The parametric fluorescence spectra, from each PPLN waveguide channel, coincide at 1550~nm, for a sample temperature of 363.0$\pm0.1$~K (see Fig.~\ref{fig:source_spectrum}). The output spectrum from the sources can be tuned (see inset of Fig.~\ref{fig:source_spectrum}). Approximately 4~nm of central wavelength shift is available for each 1~K of temperature shift across the communications band. The output photon pairs are then collected using four FLDW integrated wavelength division multiplexers (WDM), operating at 1310/1550\,nm. 
The WDMs consist of pairs of coupled waveguides whose coupling is wavelength dependent. A range of couplers were fabricated and extinction ratios of 10~dB$\pm$1~dB were obtained for an interaction length of 3.35~mm and a waveguide spacing of 7~$\mu$m, comparable to previous demonstrations~\cite{Eaton2009}. 

In order to obtain indistinguishable single photons in a single temporal mode~\cite{Mosley2008}, suitable bandpass filters are employed to achieve a coherence time greater than the pulse duration of the pump laser~\cite{Aboussouan2010}.
In the heralding arm, single photons are collected using optical circulators and core scanned, laser inscribed, fibre Bragg gratings (FBG)~\cite{Williams2013}. These are set to reflect photons at 1312\,nm with a bandwidth of 85\,GHz. 
In the heralded arm, 100\,GHz filter (DiCon MTF-08) is used to transmit energy-matched paired photons at 1548\,nm, while the pump bandwidth is 300\,GHz. 
\begin{figure}[h!]
\subfloat[]{\label{fig:rate_vpower}\includegraphics[width=0.45\textwidth]{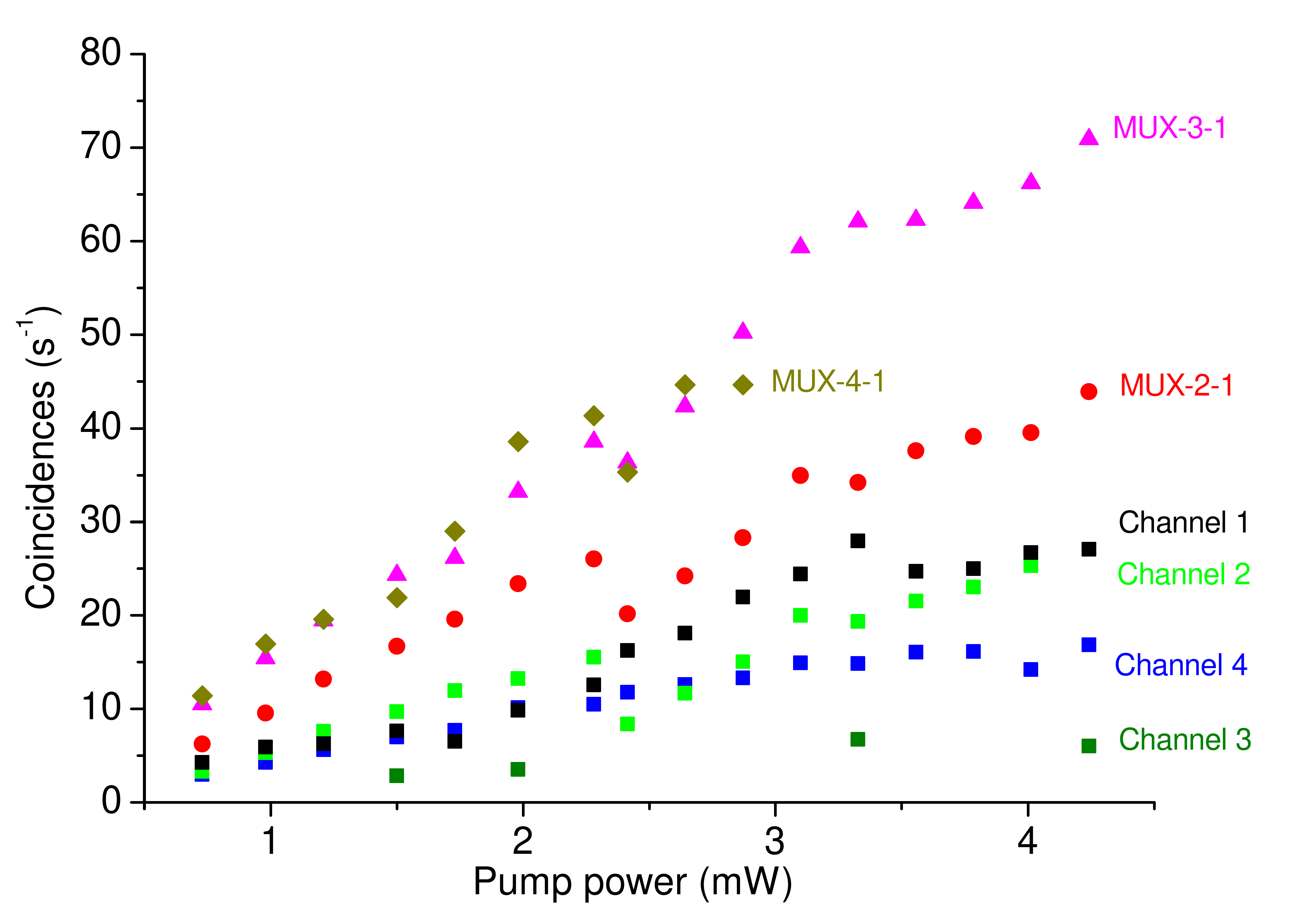}}\\
\subfloat[]{\label{fig:Car_single}\includegraphics[width=0.45\textwidth]{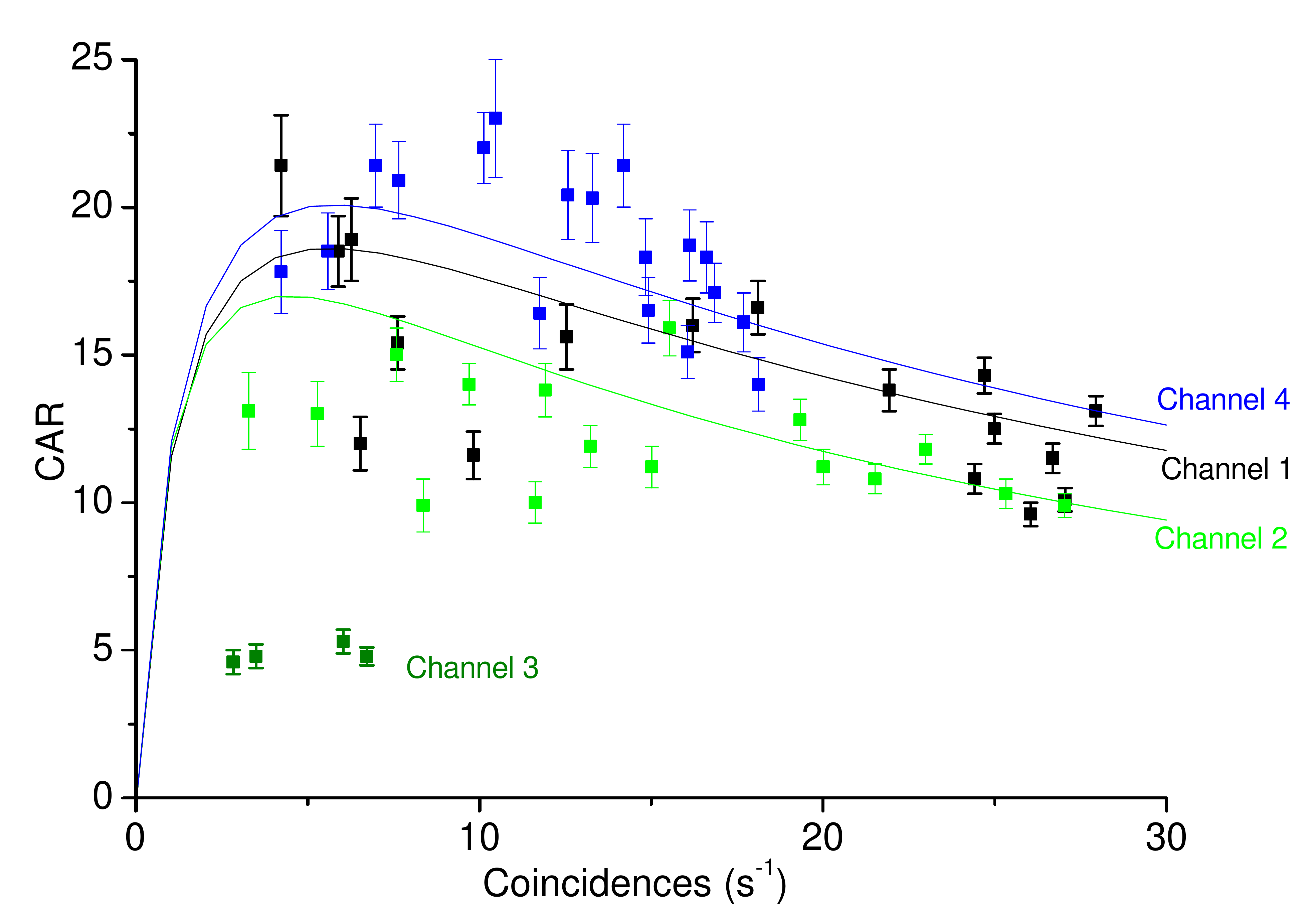}}\\ 
\subfloat[]{\label{fig:Car_MUX}\includegraphics[width=0.45\textwidth]{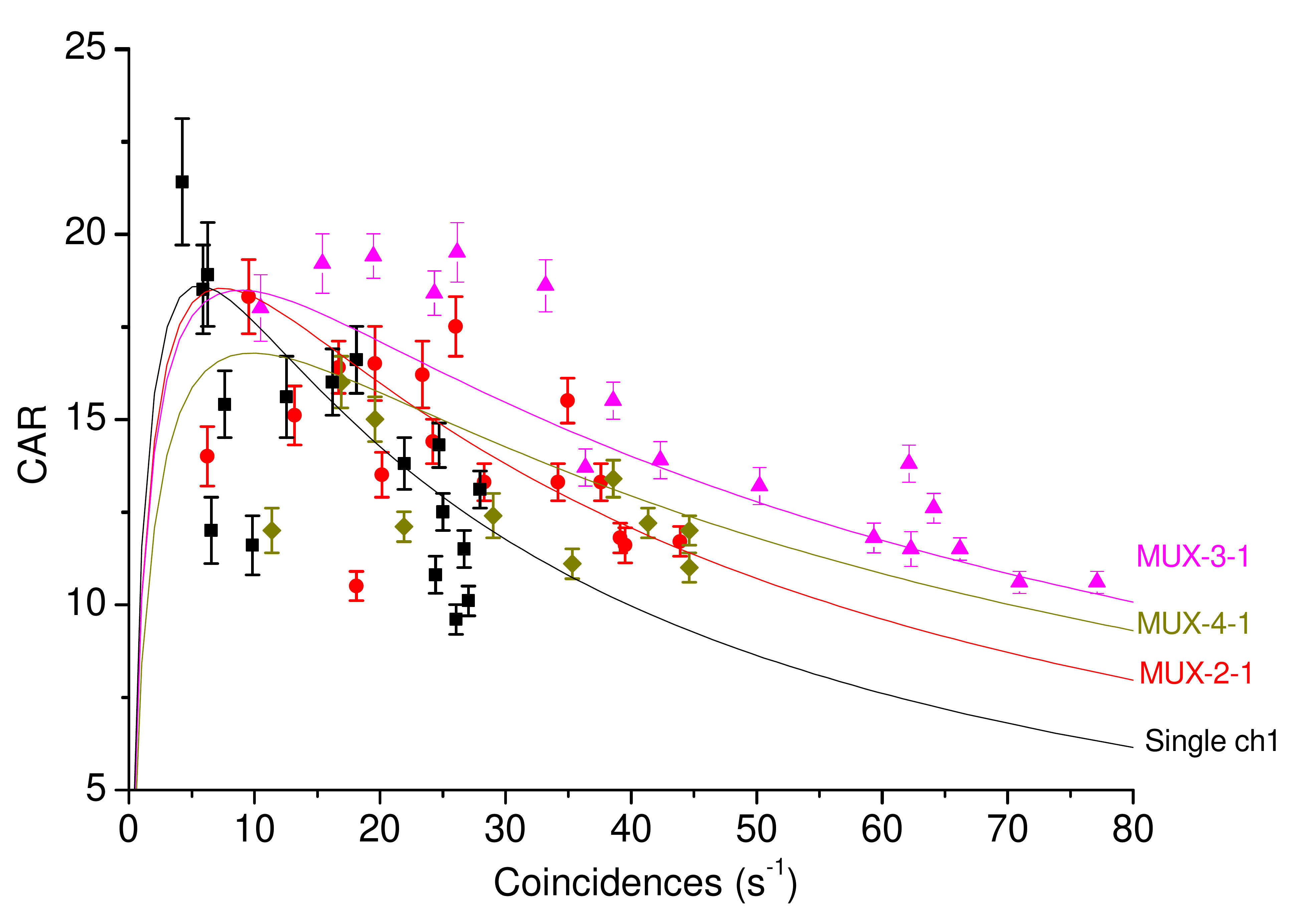}}\\ 
\caption{(a) Heralded photon rate vs. pump power for single channels and a 4-1 multiplexing scheme (b) CAR vs. coincidence rate for each single PPLN waveguide channel. (c) CAR vs. coincidences for multiplexing 2, 3 and 4 channels. Single channel 1 is shown for comparison. The solid lines in the graphs represent the predicted CAR obtained from Eq.~1 using data contained in Table 1. The error bars are given by Poissonian photon statistics.}
\end{figure}
The heralding photons are detected by four InGaAs detectors (IDQ 210), triggered by the 76~MHz pump laser, with a gate window of 5~ns and a deadtime of 3~$\mu$s. Upon detection of a photon, the output of these detectors then triggers both the lead lanthanum zirconium titanate (PLZT) switches and another InGaAs detector (IDQ 201), with an adjustable gate window. The arrangement of these switches is shown in Fig.~\ref{fig:multiplexing_config} where three switches can be used to combine the 1550\,nm outputs from each source to the final detector. Using in-fibre polarisation controllers and an analyser at the output, we ensure photons are identical in polarisation, while path lengths are controlled using variable delays to ensure maximal temporal indistinguishability.
Figure~\ref{fig:rate_vpower} shows heralded single photon rates, for individual channels and multiplexed outputs. Single channel 1 (black) and channel 2 (green) have maximum coincidence rates of 27~Hz each, for pump powers of 4.25~mW. Single channels 4 (blue) and 3 (dark green) show rates of 17~Hz and 6~Hz. By actively multiplexing, the source brightness is increased. First channel 1 and 4 (MUX-2-1) are combined using a single switch and then subsequently channels 1, 2 and 4 (MUX-3-1) are routed to a single output (See Fig.~\ref{fig:rate_vpower}). This results in coincidence rates, for all pump powers, above those obtained from any individual channel. However, when multiplexing 4 channels (MUX-4-1) the coincidence rate is approximately the same as that observed for 3 channels (MUX-3-1), a feature that is discussed in detail below. 
It can be seen in Fig.~\ref{fig:rate_vpower} that in the range of pump powers from 2.2~mW to 3.2~mW, there is a deviation from a linear trend, observed for all channels, in the source brightness. This was caused by a change in the coupling between the fibre pigtail and the 1-4 waveguide splitter. The index matching adhesive used to bind these facets can expand with temperature, due to the high laser intensity, resulting in a consistent change in output power from the 1-4 splitter. Furthermore, this adhesive has a damage threshold of approximately 20~mW, meaning that individual channels could not be pumped beyond a pump power of 4.5~mW.

The coincidence-to-accidental ratio (CAR) is a metric used to characterise the signal-to-noise ratio of a photon source. Figure~\ref{fig:Car_single} shows CAR values for all individual channels. The CAR can be expressed as a function of $C$, the coincidence rate per pulse, 
\begin{equation}
CAR= C/(\frac{C}{\eta_{S}}+d_{I})(\frac{C}{\eta_{I}}+d_{S}),
\end{equation}
where $\eta_{I,S}$ are the net signal and idler efficiencies for each channel, and $d_{I,S}$ the dark count rates per pulse~\cite{Collins2013}. It can be seen in Fig.~\ref{fig:Car_single} that the CAR values for a range of coincidences are different for each channel. Table~\ref{Table_perf} shows that this is due to differences in detector performance and intrinsic waveguide performance. Channel 3 has a low photon production efficiency, combined with a low detection efficiency and high dark count rate resulting in a CAR substantially below the other channels. 
When multiplexing, for a fixed CAR value above 10, the photon rate is increased above the single channel values while the noise remains constant (see Fig.~\ref{fig:Car_MUX}). This is the key feature of the spatially multiplexed source as it demonstrates an increase in the 
heralded single photon rate, by a factor of 3, while maintaining the corresponding signal-to-noise ratio. However, the effect of the dark count rates and low efficiency of single channel 3 reduces the CAR for the MUX-4-1, since the coincidence contribution from single channel 3 is low but its dark counts are added. When multiplexing 3 sources, excluding single channel 3, a clear increase in single photon rates can be observed over a CAR range of 10-20. We note the CAR values for lower coincidence rates do not show an enhancement due to dark count rates $>$1~kHz in the single channel heralding detectors (see Table~\ref{Table_perf}).
 
The loss in each channel also varies due to differences in individual component losses required for interfacing and multiplexing: PPLN facet extraction efficiency (3~dB), laser written WDMs (3~dB), circulators and FBGs (3~dB), pump suppression filters (1~dB), polarisation controllers (1~dB), variable delays (1.5~dB), PLZT switch (1~dB) and bandpass filter (3.5~dB). Increasing the uniformity of these individual component losses and reducing them can significantly increase the overall source performance.
The inefficiency of source 3 is mainly due to combined imperfections in the lithographic processes used for both the waveguide and periodic poling fabrication steps.
The stability of the system can be improved by using index matching adhesives showing a reduced sensitivity to thermal fluctuations which stands as the limiting factor to the temporal stability. Ideally, the 1548~nm filter bandwidth should equate to the sum of the pump and the 1312\,nm filter bandwidth, corrected by the phase-matching acceptance of the nonlinear waveguide. Any deviation from this results in a loss and reduction of the overall CAR and this factor has been estimated to be 0.5 in our case. By using a 200\,GHz filter for the 1548~nm photons that factor could be improved.

We have interfaced PPLN waveguide based sources of photons with laser written waveguide circuits to produce four spatially separated heralded photons, tunable across the communications band of 1520-1580~nm. We show the feasibility of actively routing photons to pairs of outputs and a single output using electronically controlled switches. Our output photon rates are an order of magnitude higher than a previous integrated spatial multiplexing demonstration~\cite{Collins2013} and almost double the rate of a previous demonstration of integrated bidirectional multiplexing~\cite{Xiong2013}. We have interfaced, simultaneously, the largest number of SPDC sources on a single chip which is comparable to bulk optic demonstrations~\cite{Yao2012}. This shows the practicality of hybrid integration for the preparation of single and multiple photon states. By increasing photon rates, while maintaining a fixed signal-to-noise ratio through multiplexing, we have demonstrated the scalability of our technique. 

\begin{table}[ht]
\centering
\caption{Maximum measured CAR and calculated photon pair production rate excluding all loss, $\mu$, from each PPLN channel, losses for idler and signal arms ($\eta_{I,S}$), heralding detector efficiency ($\eta_{\textrm{Detector}}$) and dark count rates ($d_{\textrm{Herald}}$) are shown for the multiplexing detection arrangement.\label{Table_perf}}
\begin{tabular}{@{}lcccc@{}}
\hline
Single channel & 1 & 2 & 3 & 4\\
\hline
$\mu$ ($\times10^{-3}$) & 12.8 & 23.1 & 1.9 & 10.8 \\
Max. CAR & 21 & 15 & 7 & 25 \\
\hline
$\eta_I$ (dB incl detector) & 19 & 21 & 26 & 20.5\\
$\eta_S$ (dB incl detector) & 33 & 33.5 & 32 & 33.1\\
$\eta_{\textrm{Detector}}$ ($\%$) & 17.5 & 17.5 & 7.5 & 12.5 \\
$d_{\textrm{Herald}}$ (kHz) & 1.8 & 1.5 & 2 & 1 \\
\hline
\end{tabular}
\end{table}

\noindent
\textit{Acknowledgement}
This research was conducted in part by the Australian Research Council Centre of Excellence for Ultrahigh bandwidth Devices for Optical Systems (CE110001018) and the Optofab node of the Australian National Fabrication Facility. 
We acknowledge financial support from the CNRS, the European ICT-2009.8.0 FET open project "QUANTIP" (grant 244026), the MARA, the Conseil R??gional PACA, the Australian Research Council DECRA program (DE130101148) and Chunle Xiong for detection apparatus from the DECRA program (DE120100226).

\end{document}